# Optically tunable Mie-resonance VO₂ nanoantennas for metasurfaces in the visible

Peter Kepič[†,‡], Filip Ligmajer[†,‡,*], Martin Hrtoň[†,‡], Haoran Ren[§], Leonardo de Souza Menezes[§], Stefan A. Maier[§,‖], Tomáš Šikola[†,‡]

† Central European Institute of Technology, Brno University of Technology, 612 00 Brno, Czech Republic

‡ Institute of Physical Engineering, Faculty of Mechanical Engineering, Brno University of Technology, 616 69 Brno, Czech Republic

§ Chair in Hybrid Nanosystems, Nanoinstitute Munich, Faculty of Physics, Ludwig-Maximilians-Universität München, 80539 München, Germany

‖ Department of Physics, Imperial College London, London SW7 2AZ, United Kingdom

* E-mail: filip.ligmajer@ceitec.vutbr.cz

## Abstract

Metasurfaces are ultrathin nanostructured surfaces that can allow arbitrary manipulation of light. Implementing dynamic tunability into their design could allow the optical functions of metasurfaces to be rapidly modified at will. The most pronounced and robust tunability of optical properties is provided by phase-change materials such as vanadium dioxide (VO₂) and germanium antimony telluride (GST), but their implementations have been limited only to near-infrared wavelengths. Here, we demonstrate that VO₂ nanoantennas with widely tunable Mie resonances can be utilized for designing tunable metasurfaces in the visible range. In contrast to the dielectric-metallic phase transition-induced tunability in previous demonstrations, we show that dielectric Mie resonances in VO₂ nanoantennas offer remarkable scattering and extinction modulation depths (5–8 dB and 1–3 dB, respectively) for tunability in the visible. Moreover, these strong resonances are optically switchable using a continuous-wave laser. Our results establish VO₂ nanostructures as low-loss building blocks of optically tunable metasurfaces.

## Introduction

Altering light propagation and its properties with increasingly better control and efficiency has definitely become a key research direction in nanophotonics. This ambitious goal is nowadays mostly achieved by metasurfaces, *i.e.*, large arrays of subwavelength metallic or dielectric nanostructures[1,2]. Ultrathin metasurface analogs of conventional optical components include lenses, waveplates, holograms, biosensors, imagers, and optical switches, which exhibit superior optical properties such as high efficiency[3,4], high bandwidth[5–7], high sensitivity[8], aberration-free functionality[9–12], and ultrafast light-modulation capacity[13]. The natural next step in the development of metasurfaces is incorporating tunability into their design[14,15]. Depending on the desired application, the metasurface function could be then gradually tuned or abruptly switched by external stimuli including applied voltage[16], current[17], or incident light intensity[18]. There are several material platforms that can endow metasurfaces with tunable properties: graphene[19,20], transition metal dichalcogenides[16,21], transparent conductive oxides[22], liquid crystals[23,24], or even bulk single chemical elements such as silicon[25] and magnesium[26]. However, the low swiching contrasts achievable with these materials or difficulties in their integration significantly limit their use in metasurfaces. On the other hand, vanadium dioxide (VO₂)[27] and germanium antimony telluride (GST)[28] stand out as two unique materials which can undergo a phase transition from a low-temperature dielectric phase to a high-temperature metallic phase. The physical mechanisms behind the $VO_2$ and GST phase transitions are different[29,30], but the associated large changes in the conductivity and dielectric function over wide frequency ranges render these two materials very promising for applications in tunable nanophotonics. While GST is mostly sought after for its non-volatility[31], $VO_2$ receives special interest due to its low transition temperature (around 67°C), low thermal hysteresis (5−10 K), and ability to withstand millions of switching cycles without degradation[32−37]. The refractive-index contrast of $VO_2$ in the visible range ($\Delta n \approx 1$) also vastly exceeds other phase-change materials.[38]

The most straightforward application of a phase-change material as a tunable metasurface is its nanostructuring using nanofabrication techniques[39−42]. Then, the resulting optical function of the metasurface can be controlled directly, while reducing delays and inefficiencies caused by the inherent material heterogeneity[43]. Although one can find a handful of works where nanostructured $VO_2$ has been considered as a nanophotonic building block in the infrared range[42,44−46], the detailed investigation of low-loss Mie resonances[47−49] in the dielectric phase of a $VO_2$ nanoantenna is still missing and its immense potential in the context of dielectric metasurfaces in the visible range remains largely untapped.

Here, we present a systematic study devoted to $VO_2$ metasurface building blocks in the visible spectral region. More specifically, we focus on the nature of Mie resonances in dielectric $VO_2$ nanodiscs and explore the possibility of their optical switching and tuning capabilities using a continuous-wave visible laser. We show that the $VO_2$ nanodiscs exhibit strong scattering resonances not only in their high-temperature plasmonic phase, but already in their low-temperature dielectric $VO_2$ phase. We experimentally investigate these resonances and elucidate their nature using analytical multipole decomposition. We quantify the switching contrast between the two $VO_2$ phases and demonstrate spectrally broad (hundreds of nanometers) and very large modulation depths (up to 8 dB in scattering and 2.5 dB in extinction). Moreover, we investigate how these resonances get sharper when single nanodiscs form an extended array, which is also complemented by spectral sharpening of the extinction modulation depth. We show that these sharp array modulation depths are strongly dependent on the nanodisc size and can thus be used for creation of metasurfaces with spatially variable modulation depths, a feature useful for tunable hologram generation or anti-counterfeiting applications. Finally, we also test optical switching and tuning capabilities of the $VO_2$ nanodiscs for their potential use in optically tunable metasurfaces. Namely, using a continuous-wave visible laser, we were able to drive the Mie resonances through the visible range with narrow hysteresis and extinction modulation depth of 1.5 dB. Altogether, our results prove the immense potential of dielectric $VO_2$ nanodiscs for design of optically tunable visible metasurfaces.

## Single $VO_2$ nanodiscs

Functional properties of any metasurface are given by a complex interplay between the effects of the individual building blocks and by the collective effects arising from their periodic arrangement. We will therefore first focus our attention on the properties of single $VO_2$ nanodiscs, and the collective lattice effects will be discussed later in the text.

In Figure 1, we show an overview of simulated scattering cross-sections of single $VO_2$ nanodiscs with various diameters ranging from 100 nm to 300 nm for both the dielectric (30°C) and metallic (90°C) phases of the material. The well-known and often-used plasmonic resonances of the metallic $VO_2$ are apparent in the near-infrared (NIR) range[45,50], and they disappear when VO₂ goes to its dielectric phase (i.e., at lower temperatures). In the visible (VIS) region, however, an additional set of strong scattering resonances appears in both material phases of VO₂. Like their NIR plasmonic counterparts, these resonances are strongly dependent on size — red-shifting across the whole VIS range as the nanodisc diameter grows. We identify these scattering peaks as Mie resonances, considering positive magnitudes of the real parts of the complex dielectric function ($\varepsilon = \varepsilon_1 + i\varepsilon_2$) of VO₂ (see the ellipsometry data in Figure S1 in Supporting Information). In the NIR region, the $\varepsilon_1$ of the low-temperature (dielectric phase) VO₂ is positive as well but nanodiscs are too small to exhibit any resonances in this spectral range. The nature of these resonances will be investigated in greater detail later below. When the temperature increases and VO₂ transforms into its metallic phase, the $\varepsilon_1$ drops below zero in the NIR region, and the localized plasmonic resonances in the nanodiscs emerge. In the VIS, the situation is different, however: The $\varepsilon_1$ is always positive and VO₂ behaves as a dielectric, regardless of its material phase. The common notion of "metallic" phase for high-temperature VO₂ (based on its infrared and static response) is thus a sort of misnomer at VIS wavelengths.

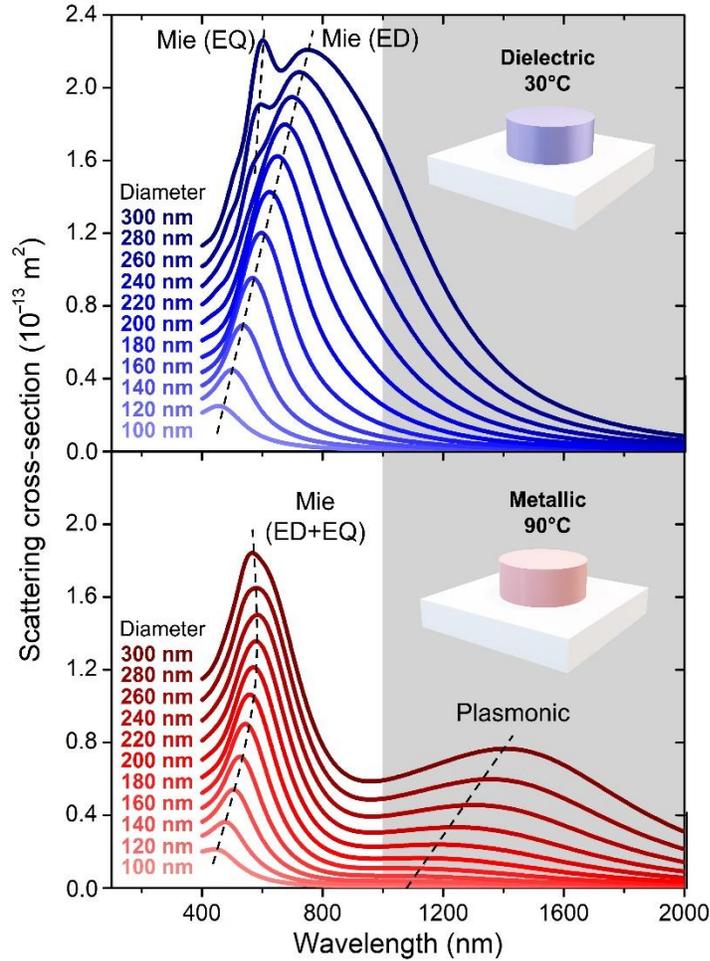

***Figure 1**: Simulated scattering cross-sections of dielectric (top) and metallic (bottom) single VO₂ nanodiscs with a varying diameter and fixed height (200 nm) on a fused silica substrate. The dashed lines track the positions of Mie (electric dipole, ED; electric quadrupole, EQ) and plasmonic resonances. The gray area denotes the region of plasmonic VO₂, often used as a tunable medium in metasurfaces.*

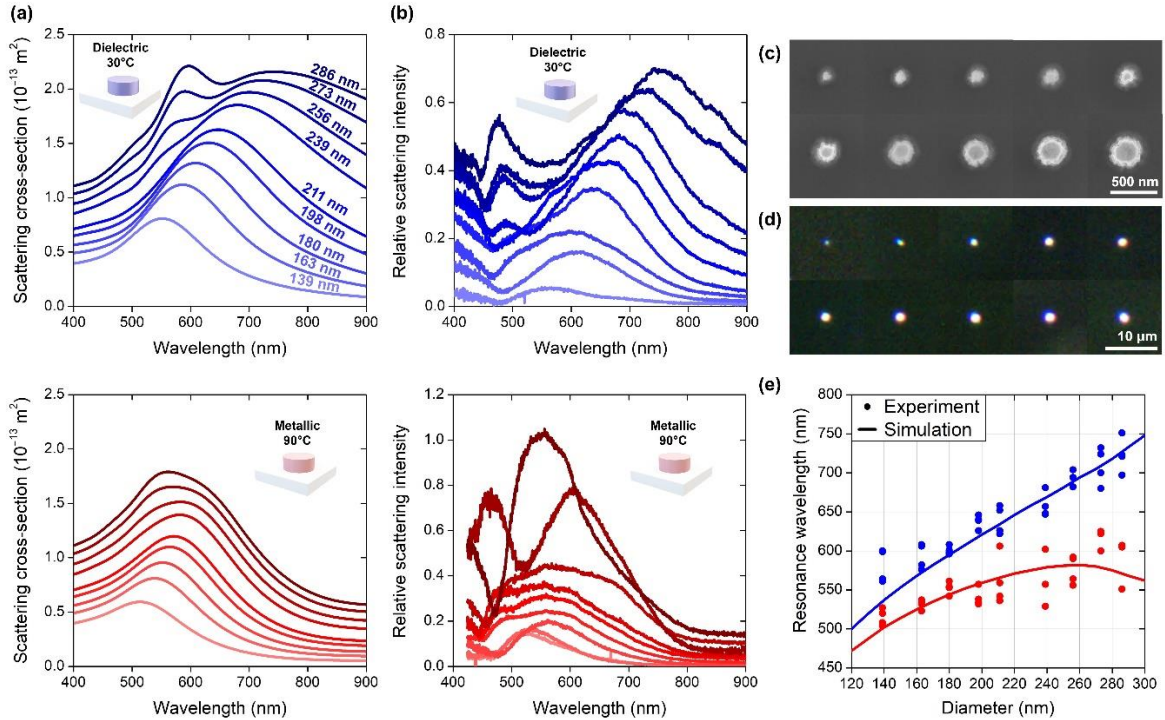

***Figure 2***: *(a) Simulated scattering cross-sections and (b) experimental dark-field optical scattering spectra for single VO₂ nanodiscs (height 200 nm, diameters as measured from SEM images) on fused silica substrate in the dielectric (top) and metallic (bottom) VO₂ phases. (c) SEM and (d) dark-field optical microscopy images of the measured VO₂ nanodiscs. Note that the scattering intensity from the smallest nanodisc (nominal diameter 120 nm) is not reported due to excesive noise in the spectra. (e) Resonant wavelengths extracted from the measured (dots) and simulated (lines) data from part (a,b). For each target diameter we have measured three to four different nanodiscs to partially alleviate the effect of fabrication imperfections and thermal instabilities. Blue and red color correspond to the dielectric and metallic VO₂ phases, respectively.*

Note that we report here only the results for nanodiscs of 200 nm height, but a similar behavior can be observed with lower (100 nm) and higher (300 nm) nanodiscs, as shown in Figure S2 in Supporting Information.

To experimentally test these new theoretical predictions, we fabricated spatially isolated VO₂ nanodiscs with nominal diameters ranging from 140 nm to 290 nm using electron beam lithography (see Methods). Scanning electron microscopy (SEM) and dark-field (DF) optical microscopy images of the resulting nanodiscs are shown in Figure 2c,d. We measured single particle scattering spectra using DF optical spectroscopy (Figure 2b), where we indeed observed the predicted strong scattering resonances. Note that for some of the largest nanodiscs, double-peak resonances appeared, which we ascribe to the slight non-normal excitation and detection related to the finite numerical aperture of the used microscope objectives. For easier comparison of the spectral positions and lineshapes of the dominant resonances, we rerun numerical simulations from Figure 1, now with the dimensions matching the fabricated nanodiscs (Figure 2a). In Figure 2e, we then compare the resonance wavelengths extracted from the measurements with those from the new simulations. Note that for each target diameter we have measured three to four different nanodiscs to partially mitigate the adverse effect of fabrication imperfections and thermal instabilities. To sum up, these results show VO₂ is a proper tunable material for nanoantennas, which can be used not only in its plasmonic phase, but also in its dielectric phase due to strong Mie resonances spanning a large portion of the VIS range.

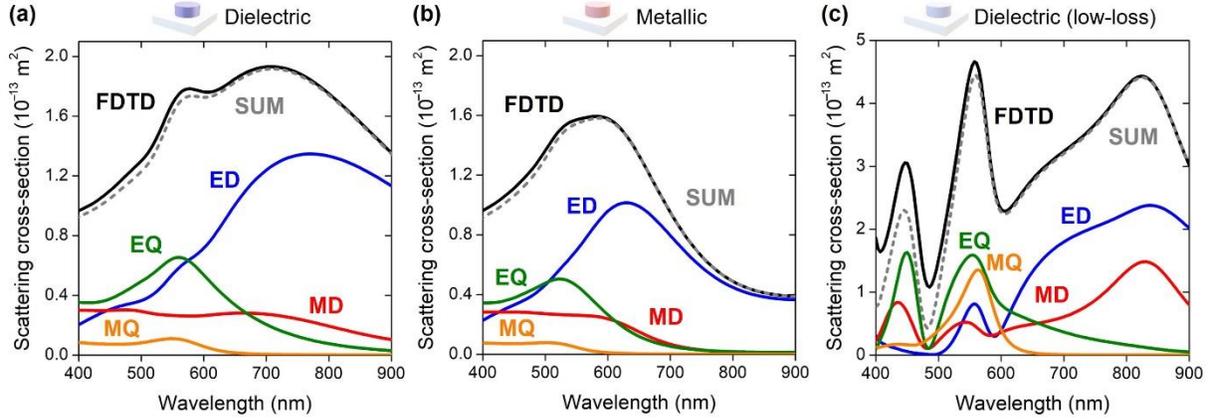

**Figure 3**: *Multipole scattering decomposition of a single dielectric (a) and metallic (b) VO₂ nanodisc with 200 nm height and 270 nm diameter on a fused silica substrate. The "sum" (grey dashed line) represents a sum of the four multipoles (color lines) and of their cross-terms that do not cancel out in the presence of a substrate. The total scattering cross-section (black line) calculated directly in the FDTD solver is plotted for comparison. (c) Multipole decomposition for the same single dielectric VO₂ nanodisc when its extinction coefficient (k) is artificially set to zero and its refractive index (n) is flattened to the mean value of 2.89.*

To properly understand the nature of the observed Mie resonances in the VIS region, we performed analytical multipolar decomposition of the scattered fields for a single model VO₂ nanodisc (diameter 270 nm, height 200 nm). Using a modified decomposition approach involving substrate effects (see Methods), we broke down the simulated all-directional scattering cross-section into contributions from electric and magnetic dipoles and quadrupoles (Figure 3a,b). For the low-temperature dielectric VO₂ phase, we can see that a combination of electric dipole (ED) and electric quadrupole (EQ) modes dominates the scattering spectrum. The same notion holds also for the Mie resonances in the high-temperature VO₂ phase — the resonances just get slightly weaker and narrower, due to the lower index of refraction and lower losses of the VO₂ in the VIS region (Figure S1). Magnetic dipole (MD) modes, which are indispensable in many modern applications of dielectric nanostructures[51], are very weak in both VO₂ phases. The absence of strong MD modes is caused by the nonzero absorption coefficient of VO₂ related to its polycrystallinity[52–55], imperfect stoichiometry[56–59], and mostly to the specific band structure of the strongly correlated material[60–63].

To shed more light on the role of losses in this situation, we recalculated the multipolar decomposition of the same VO₂ nanodisc, while we artificially set the extinction coefficient $k$ (*i.e.*, the imaginary part of the complex refractive index) of VO₂ to zero and flattened the real part of its refractive index $n$ to the mean value of 2.89 (see Figure 3c). Note that although the low values of the VO₂'s extinction coefficient ($k \sim 0.1$) are not common, they are attainable by carefully optimizing the VO₂ deposition technique.[64,65] This loss alleviating modification results in significantly enhanced magnetic multipoles in the dielectric VO₂ nanodisc. Especially the MQ gets almost as strong as its electric counterpart. The overlapping electric and magnetic multipoles are indispensable for building so-called Huygens metasurfaces, which exhibit a significantly reduced back-scattering response[66,67]. We demonstrate this effect in the context of VO₂ nanodiscs by numerical simulations presented in the Supporting Information (Figure S3), where a single artificially lossless $VO_2$ nanodisc exhibits the forward-to-backward scattering ratio >10. We can therefore conclude that if a carefully optimized low-loss $VO_2$ deposition technique is used one could significantly enhance displacement currents in $VO_2$ nanostructures, which would then act like Huygens' sources or toroidal multipole sources[68,69] — in the same way as conventional dielectrics like silicon[70,71], but with the added unique benefit of tunability.

## VO₂ nanodisc arrays

The knowledge of scattering properties of single nanostructures is essential for understanding the general behavior of any metasurface composed of these tunable building blocks. It is well known, however, that when such strongly scattering nanostructures are arranged in a periodic lattice, their collective optical response can be significantly modified by both near- and far-field interactions amongst them[72–75]. We demonstrate this effect in the context of our $VO_2$ nanodisc system by comparing calculated extinction cross-sections of a single nanodisc with those of an array of 13×13 interacting nanodiscs, as shown in Figure 4a. From this comparison it is obvious how interactions within the array lead to spectral shifts and, most importantly, to sharpening of the resonances in both $VO_2$ phases. To experimentally investigate this feature, we fabricated also arrays of $VO_2$ nanodiscs with the same height (200 nm) and variable diameters $D$ as before (array size = 100 μm × 100 μm, pitch = 1.5$D$). Instead of dark-field spectroscopy, for which the numerical simulations of large arrays get problematic due to large non-normal angles of incidence[76], we used bright-field transmission spectroscopy as a proxy for extinction spectra evaluation (see Methods)[77]. The results are shown in Figure 4b, where we can indeed observe the expected lattice effect in action. For arrays of small nanodiscs, the dominant dipolar resonances get sharper, while for the arrays of larger nanodiscs, their higher order multipoles become more pronounced and separated (cf. Figure 2a,b; see also multipole expansion including the array effect in Figure S5 in Supporting Information). This array-induced sharpening can be also clearly visualized and quantified by respective full widths at half maximum (FWHM) of the dipolar resonances, which we report in Supporting Information (Figure S4a). The sharpening of the resonances can be explained by array-imposed restrictions on the spatial structure of the scattered far field and subsequent decrease of the number of radiative decay channels.[78]

Although the resonance bandwidth is an important metric, the switching contrast between the two $VO_2$ phases is even more important in the realm of tunable metasurfaces. The switching contrast is often characterized by the so-called modulation depth[79], obtained from the measured scattering or extinction intensities corresponding to the two $VO_2$ phases ($10\log(I_{dielectric} / I_{metal})$). In Figure 4c, we first return back to the single-vs-array discussion and plot the simulated modulation depths extracted from Figure 4a. One thing worth noticing is the modulation depth magnitude, which exceeds 2.5 dB at resonance —— a value generally considered challenging to achieve with state-of-the-art tunable metasurface building blocks[80]. It is also obvious from Figure 4a how interactions within an array again lead to sharpening and increase of the modulation depth when compared with the single-nanodisc situation. But unlike with the resonances themselves, the effect is not very strong. A more interesting behavior can be observed when we inspect how the modulation-depth spectra depend on the nanodisc size: Figure 4d shows how the resonantly enhanced modulation depth closely tracks the spectral positions of the $VO_2$ nanodisc resonances in our experiments. The nanodisc size can thus directly encode the modulation depth at a selected wavelength. This property can be used for designing metasurfaces with spatially varying extinction changes of units of decibels associated with the externally controlled VO₂ material phase. Moreover, when only the scattering contribution is considered as it is necessary in the context of dark-field applications, the modulation depths reach even larger values of 5–8 dB at the respective resonances (Figure S4b in Supporting Information), thus offering extremely high switching contrasts. To conclude this section, we examined in detail how the interaction within nanodisc arrays influences their collective response and that it can be harnessed to yield sharp resonances or, conversely, to mitigate them when a broadband response is required. We also quantified the tunable capacity of the VO₂ nanoantennas, demonstrating very large modulation depths in units of decibels. We believe these results set VO₂ at the forefront in terms of the switching contrast in the visible, with potential for further improvements towards even more demanding applications.

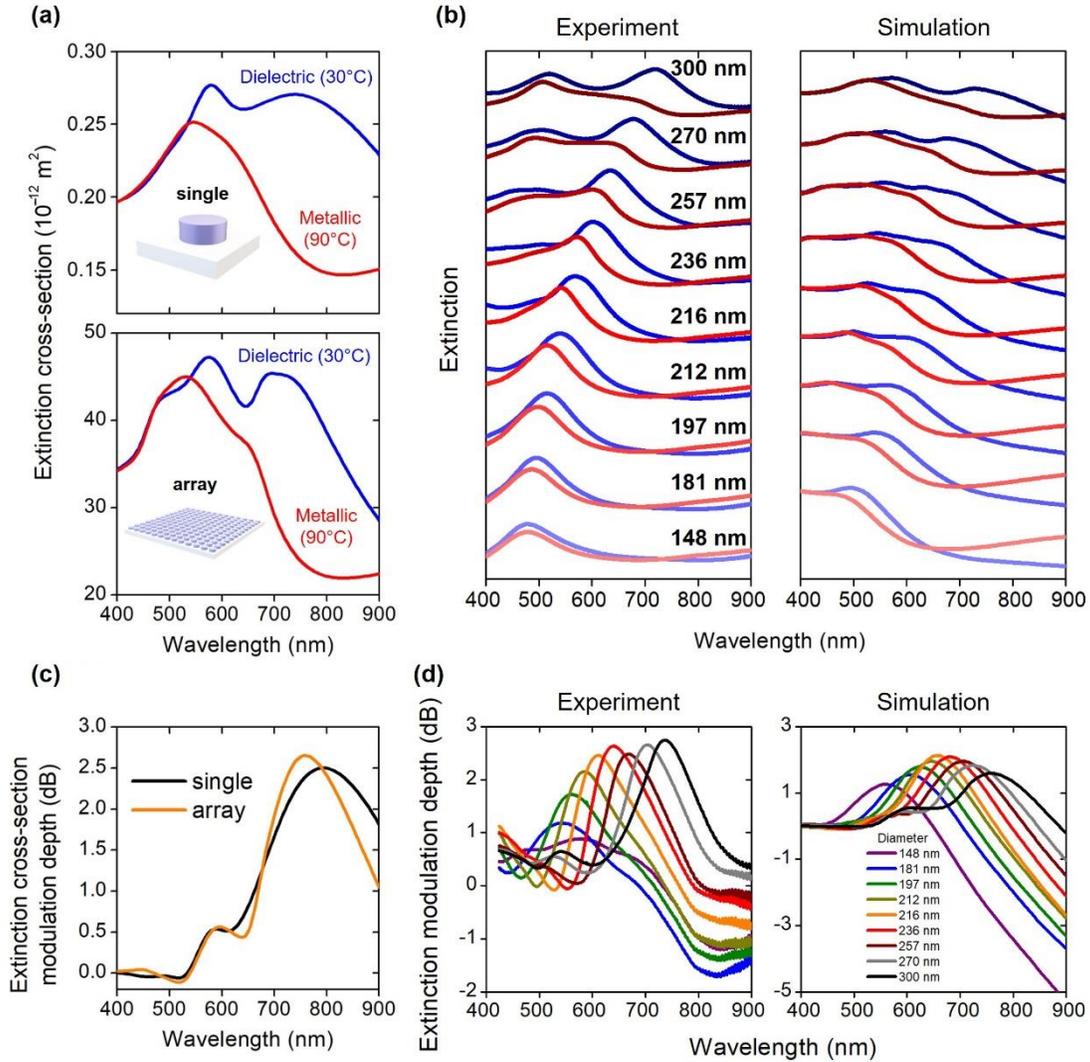

***Figure 4***: *(a) Calculated extinction cross-sections of a single nanodisc (top) and of a finite (13×13) nanodisc array (bottom) with height h = 200 nm, diameter D = 270 nm and pitch P = 1.5D. (b) Measured and simulated extinction spectra of dielectric and metallic VO₂ nanodisc arrays with the same height and pitch, and varying diameters (as labeled in the graph). (c) Modulation depths corresponding to the VO₂ phase change, defined as $10log(I_{dielectric}/I_{metal})$, calculated directly from the extinction cross-section spectra in (a). (d) Extinction modulation depth spectra calculated from data in (b). Note that the spectra were smoothed by an FFT filter with a 50-point window to limit the noise.*

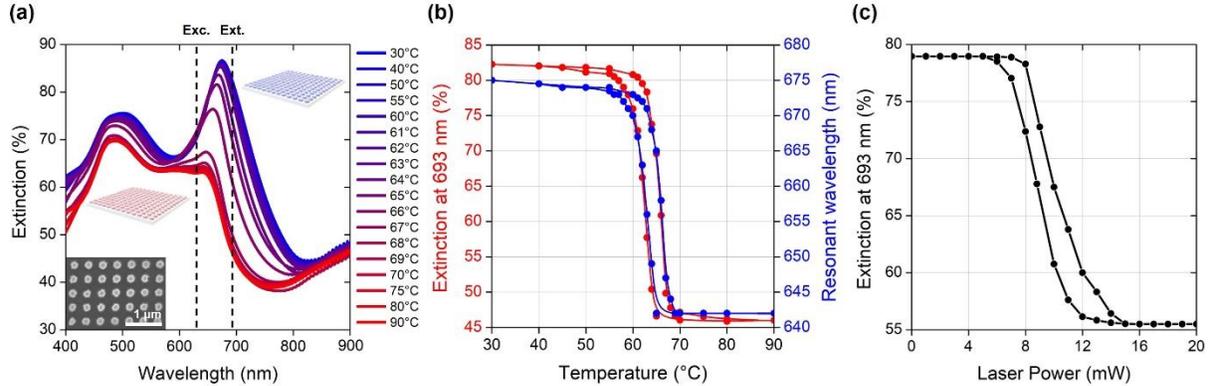

**Figure 5**: *(a) Experimental extinction spectra of a VO₂ nanodisc array (270 nm nanodisc diameter) during the heating cycle, which is driving the VO₂ through its phase transition. The inset is a SEM micrograph of the array. The dashed vertical lines label the excitation (Exc.) and extinction measurement (Ext.) wavelengths in (b) and (c). (b) Hysteresis-like behavior of the measured extinction at 693 nm (left axis) and of the resonant wavelength of the dipolar resonance (right axis), as extracted from (a) as a function of temperature. (c) Extinction of the same VO₂ nanodisc array measured at 693 nm upon continuous illumination by a continuous wave laser (633 nm) with gradually increasing power.*

## Optically induced tuning of the resonances

To demonstrate suitability of the VO₂ nanoantennas as building blocks for optically tunable metasurfaces in the visible range, we selected a representative array of nanodiscs (270 nm in diameter) and performed laser-induced optical tuning of its optical response. For calibration, we first measured the extinction spectra as a function of temperature, while utilizing a resistive heater stage (see Figure 5a). In between the two extreme VO₂ phases discussed above, we observed the characteristic gradual shift of the spectra with hysteresis behavior during the heating-cooling cycle. This extinction change was most pronounced between the two dipolar resonances of the two VO₂ phases around the wavelength of 700 nm. For better visualization of these large changes, we extracted extinction at 693 nm during the heating cycle as a function of temperature into Figure 5b, together with the temperature-dependent spectral position of the dipolar resonance peak. The peak shift of 33 nm, which is accompanied by considerable resonance damping, results in the extinction modulation depth of 2.5 dB. Note that an unstructured 200 nm thin VO₂ layer from the same batch exhibited a modulation depth at this wavelength of mere −0.1 dB, thus illustrating once again how strong the influence of nanostructuring and Mie resonances is. Finally, the demonstration of light-triggered phase transition tuning of the VO₂ nanodisc arrays was performed using a 633 nm continuous wave laser. In Figure 5c we show how by illuminating the array with laser light and by varying its intensity, we were able to gradually drive the nanodiscs through the VO₂ phase transition and observe laser-triggered extinction modulation of the visible Mie resonances with a modulation depth of 1.5 dB. This demonstration confirms that optical tuning of the visible Mie resonances with visible light is indeed possible and paves the way for optically controlled tunable metasurfaces. In applications where a continuous flow of energy to the substrate is unwanted, ultrashort phase transition of VO₂ through primarily electronic excitation[29,81] or electron injection[82,83] is possible, pushing the switching cycles towards picosecond timescales.

## Conclusion

In conclusion, we have fabricated and systematically characterized VO$_2$ nanodiscs with a special focus on their Mie resonances. In previous studies devoted to optical properties of VO$_2$ nanostructures, the emphasis was solely on thermal switching between low- and high-temperature VO$_2$ phases to tune near-infrared plasmonic resonances. In other words, the focus was only on switching the near-infrared plasmonic state on and off, while the visible range was disregarded due to the absence of plasmonic properties. Here we show that in the visible range, both material phases of VO$_2$ behave like a lossy dielectric and the Mie resonances morph from one type to the other during the phase change. Using multipole decomposition analysis, we elucidated the nature of these Mie resonances and speculated about high-quality resonances achievable with low-loss VO$_2$. We also studied how the Mie resonances get sharper and more pronounced when the VO$_2$ nanodiscs are arranged into larger arrays. We quantitatively compared the tunable properties of the nanodiscs using a spectrally resolved metric of the modulation depth and showed how the strong and broadband modulation of single nanodisc scattering (5–8 dB) contrasts with sharp yet more modest modulation of nanodisc array extinction (1–3 dB). We envision that the size-dependent modulation depths provided by VO$_2$ nanodiscs could be used for creation of metasurfaces with spatially encoded modulation depths for tunable holograms or anti-counterfeiting applications. Finally, we have also investigated the potential of these nanostructures as building blocks for optically tunable metasurfaces in the visible range. We evaluate how the nanostructuring leads to much higher modulation depths compared to a bare unstructured film (2.5 dB vs 0.1 dB) and how the illumination by a continuous wave laser leads to gradual tuning of the Mie resonances across the visible range. Together, all these experiments and simulations verify that VO$_2$ nanostructures can be used as tunable elements for Mie or Huygens metasurfaces in the visible range, on a material platform that supports ultrafast light switching on picosecond timescales. One can also envision active spatial programming of metasurfaces built from these building blocks in a serial[84] or a parallel[85,86] manner, putting down the burden of fixed optical functions defined during fabrication of a metasurface.

## Methods

### Fabrication

VO$_2$ thin films were deposited by evaporating stoichiometric VO$_2$ powder (Mateck) in an electron beam evaporator (Bestec, 8 kV, 32 mA, 1 Å/s) at room temperature, followed by 10 minutes of post-annealing at 500°C in a vacuum furnace under 20 sccm of O$_2$. The VO$_2$ nanodiscs were fabricated by electron beam lithography, using a 300 nm thick positive resist (Allresist, AR-P 6200.13) on fused silica substrates. The desired patterns were exposed using the Tescan MIRA3 electron beam writer at 30 kV, 27 pA, 0.005 μm step size and 230 μC/cm$^2$ dose factor. The samples were developed by amyl acetate (AR 600-546). Subsequent evaporation of a 200 nm thick VO$_2$ layer from a stoichiometric powder (Mateck) was performed using an electron beam evaporator (Bestec, 8 kV, 32 mA, 1 Å/s) at room temperature. The lift-off process was carried out with the aid of Allresist's ARP 600-71 remover. Finally, the samples were annealed at 500°C in a vacuum furnace under 20 sccm of O$_2$ for 10 minutes.

### Characterization

The complex dielectric function of the VO$_2$ thin films and their transmission spectra were characterized using a UV-NIR spectroscopic ellipsometer (J. A. Woollam, V-VASE). Dark-field scattering spectra of the single nanodiscs and transmission spectra of the nanodisc arrays were measured using a confocal microscopy system (WITec alpha300 RA, illumination by a dark-field condenser with NA=1.2, detection by a 100× NA=0.9 objective). Scattering intensity was measured as $I_{sca} = \frac{I_{nd} - I_{bg}}{I_{ref} - I_{bg}}$, where $I_{nd}$ is the light intensity collected from a single nanodisc, $I_{bg}$ is the intensity of a dark-frame background, and $I_{ref}$ is the intensity scattered from a uniform diffuse reference standard made of polished PTFE. Extinction spectra were calculated from the measured transmittance (illumination by a 60× NA=0.8 objective and detection by a 10× NA=0.25 objective) as $E = 1 - \frac{T_{nd}}{T_{ref}}$, where $T_{nd}$ is the transmittance through the nanodisc array and $T_{ref}$ refers to the transmittance through the bare fused silica substrate. For the temperature control in all experiments we used a feedback-controlled heating stage, which included a resistive heater and a thermocouple. The optical switching was performed by illuminating the samples with a 633 nm continuous wave laser

**Simulations and multipolar decomposition**

Scattering and extinction cross-sections, as well as extinction spectra, were calculated using the finite-difference time-domain (FDTD) method implemented in the Lumerical FDTD Solutions software. The size of the FDTD region varied with the number and size of the discs within the simulation. The distance between the structures and the simulation region boundary was always kept to be at least a half of the longest recorded wavelength (careful convergence testing was carried out). Conformal meshing (mesh order 4) was adopted everywhere except the VO$_2$ discs themselves, where we opted for staircase meshing with a 10 nm step. Both the single disc and disc array simulations employed a total-field scattered-field (TFSF) source with appropriate symmetry conditions and perfectly matched layers (PML) boundary conditions. The scattered power was measured with a monitor box enclosing both the structures and the source, while the extinction spectra were evaluated from the forward-scattered far-field amplitude using the optical theorem[87]. The procedure for calculation of the forward-scattered far-field amplitude involved recording of the polarization currents induced within the discs and subsequent employment of the Green's function formalism[88] that fully accounts for the presence of the glass substrate. The multipolar decompositions were carried out following the very same concept, only the polarization current was first broken down into individual multipolar contributions[89] and the forward-scattered far-field amplitude was then evaluated separately for each multipole moment. Note that in the case of disc arrays, the current distribution entering this procedure was the average polarization current induced within the discs constituting the array. Furthermore, apart from the contributions corresponding to pure multipole moments, there are also additional cross terms accounting for their mutual interactions. We refrained from plotting them to keep the respective figures lucid, but the overall response (denoted as 'SUM') incorporates them all.

# Author information

**Corresponding Authors**

* E-mail: filip.ligmajer@ceitec.vutbr.cz

**ORCID**

Peter Kepič: 0000-0002-9098-1900

Filip Ligmajer: 0000-0003-0346-4110

Martin Hrtoň: 0000-0002-3264-4025

Haoran Ren: 0000-0002-2885-875X

Leonardo de Souza Menezes: 0000-0002-8654-1953

Stefan A. Maier: 0000-0001-9704-7902

Tomáš Šikola: 0000-0003-4217-2276

**Author Contributions**

F.L. and P.K. conceived the idea and designed the study. P.K. fabricated and characterized (evaporation, annealing, EBL, SEM, optical measurements) the sample. F.L. performed the additional sample characterization (optical ellipsometry). M.H. wrote the script for multipolar decomposition. P.K., F.L. and M.H. performed the simulations and multipolar decomposition. H.R., S.M., and T.S. oversaw the overall research project. F.L, P.K., and M.H. co-wrote the manuscript and all authors commented on it.

**Notes**

The authors declare no competing financial interest.

## Acknowledgments

This work has been supported by the Grant agency of the Czech Republic (21-29468S and 20-01673S) and by Brno University of Technology (FSI-S-20-6485). We also acknowledge CzechNanoLab Research Infrastructure supported by MEYS CR (LM2018110). H. R. acknowledges funding support through a Humboldt Research Fellowship from the Alexander von Humboldt Foundation.

## Supporting Information

Measured dielectric function of the $VO_2$ thin film; Scattering cross-section spectra of single $VO_2$ nanodiscs with varying height; Forward-to-backward scattering cross-section ratio for a $VO_2$ nanodisc; Full width at half maximum of the scattering and extinction resonances; Single-nanodisc scattering modulation depths; Multipole extinction decomposition of a $VO_2$ nanodisc array.

# Supporting Information

## for

# Optically tunable Mie-resonance VO₂ nanoantennas for metasurfaces in the visible

Peter Kepič[†,‡], Filip Ligmajer[†,‡,*], Martin Hrtoň[†,‡], Haoran Ren[§], Leonardo de Souza Menezes[§], Stefan A. Maier[§,||], Tomáš Šikola[†,‡]

† Central European Institute of Technology, Brno University of Technology, 612 00 Brno, Czech Republic

‡ Institute of Physical Engineering, Faculty of Mechanical Engineering, Brno University of Technology, 616 69 Brno, Czech Republic

§ Chair in Hybrid Nanosystems, Nanoinstitute Munich, Faculty of Physics, Ludwig-Maxilimians-Universität München, 80539 München, Germany

|| Department of Physics, Imperial College London, London SW7 2AZ, United Kingdom

* E-mail: filip.ligmajer@ceitec.vutbr.cz

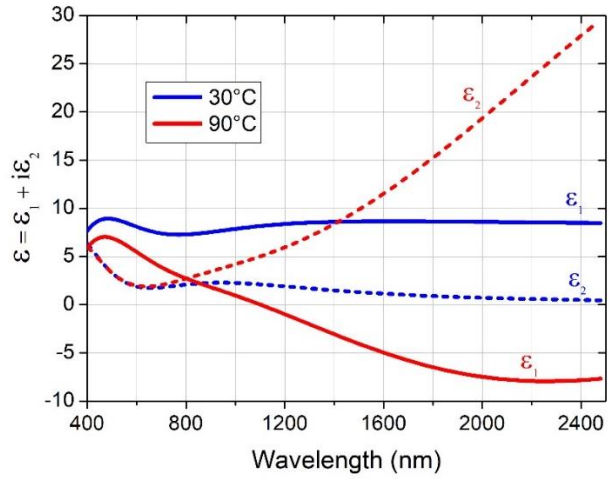

**Figure S1**. Real ($\varepsilon_1$) and imaginary ($\varepsilon_2$) parts of the dielectric function of a 93 nm thick $VO_2$ thin film on a silicon substrate, extracted from a fit to spectroscopic ellipsometry data.

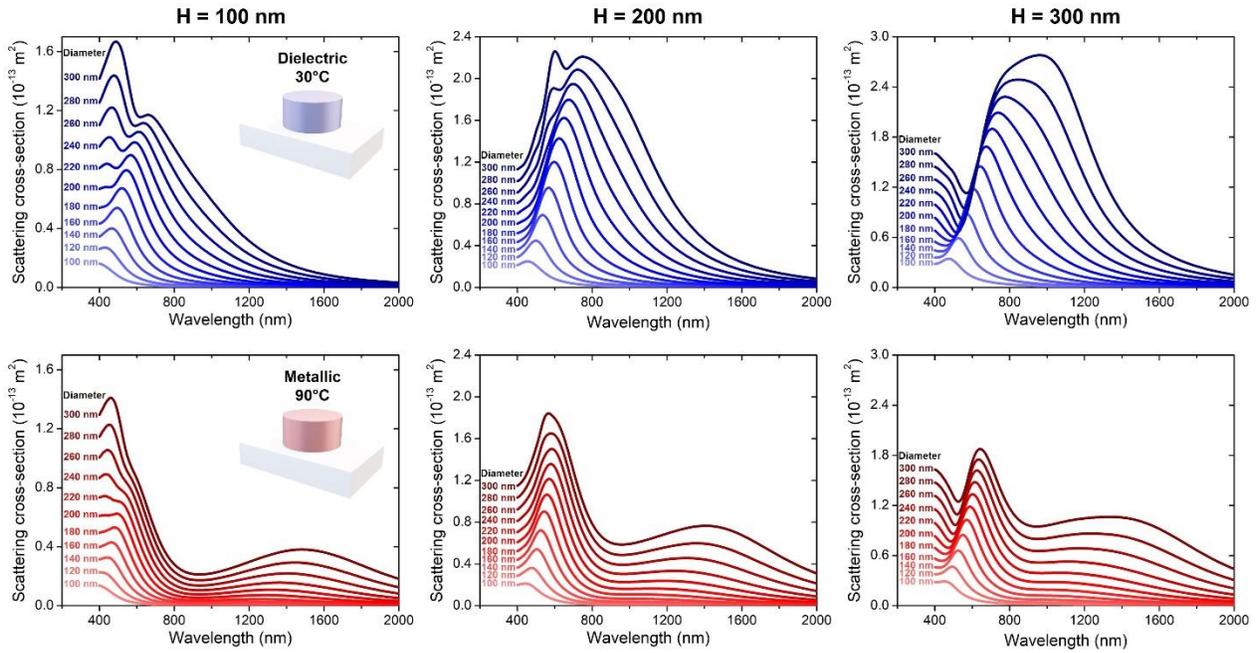

**Figure S2**. Simulated scattering cross-section spectra of single $VO_2$ nanodiscs with varying height (100 nm – 300 nm) and diameter (100 nm – 300 nm) in the dielectric phase (top) and the metallic phase (bottom).

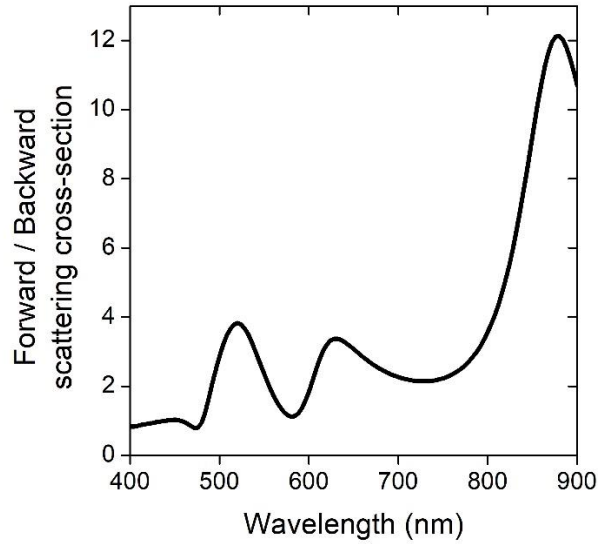

**Figure S3.** Forward-to-backward scattering cross-section ratio calculated for an array of VO$_2$ nanodiscs in the dielectric phase (nanodisc height $h$ = 200 nm, diameter $D$ = 270 nm, pitch P = 1.5D).

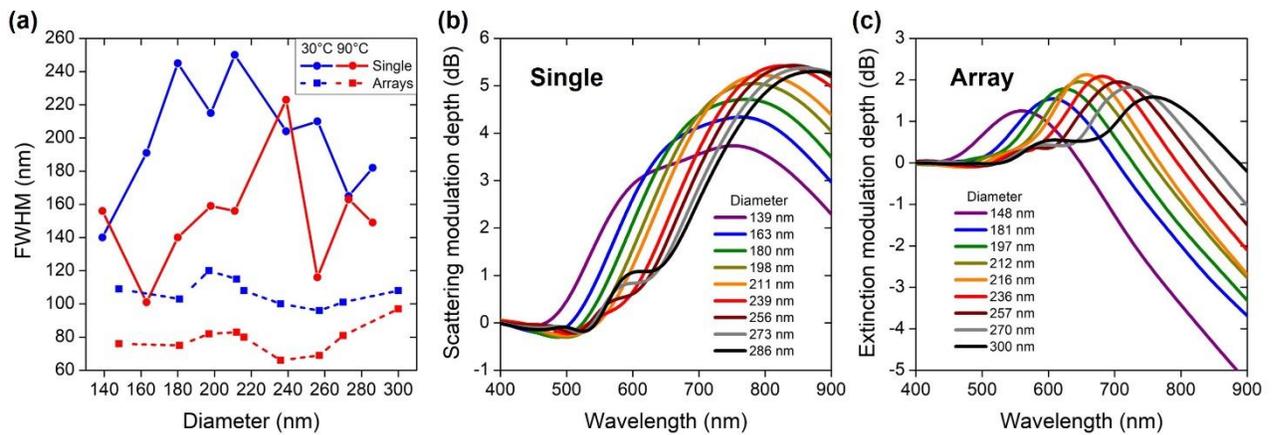

**Figure S4.** (a) Full width at half maximum (FWHM) of the scattering (single nanodiscs, solid lines) and extinction (nanodisc arrays, dashed lines) resonances, as extracted from Gaussian fits to the respective measured spectra shown in the main article. Note that when higher-order resonances became apparent in the spectra of larger nanodiscs, we used two Gaussian functions for the fit but we report here the FWHM only for the dipolar (long wavelength) resonance. (b,c) Modulation depths corresponding to the VO$_2$ phase change, defined as 10log(I$_{dielectric}$/I$_{metal}$), calculated directly from the single nanodisc scattering spectra (b) and the nanodisc array extinction spectra (c).

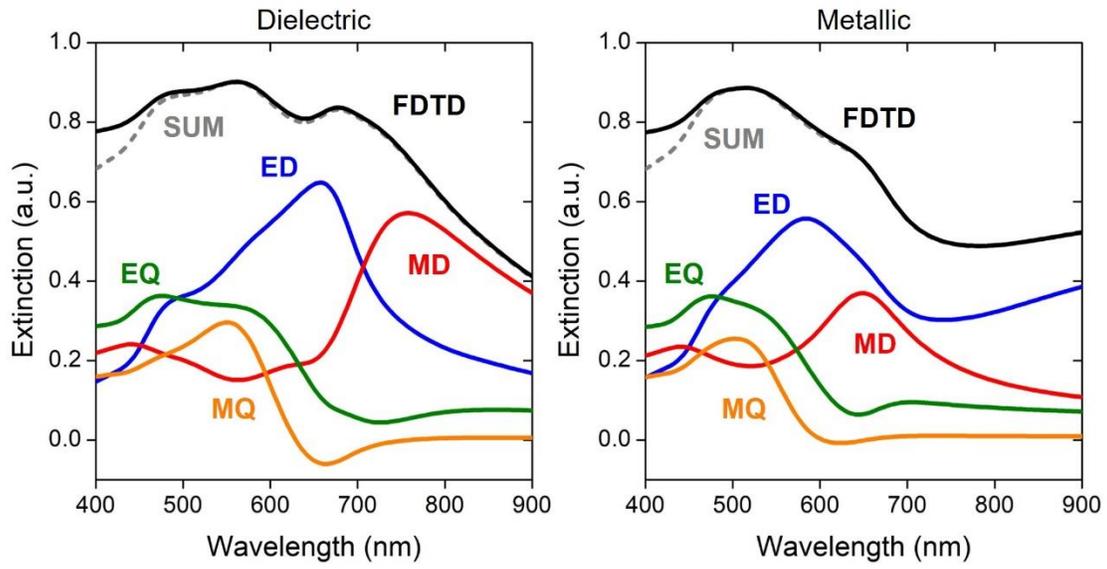

**Figure S5**. Multipole extinction decomposition of the dielectric and metallic VO₂ nanodisc array on a fused silica substrate (nanodisc height *h* = 200 nm, diameter *D* = 270 nm, pitch P = 1.5D).